\documentclass[aps,twocolumn,groupedaddress]{revtex4}
\usepackage{psfig}
\begin{document}
\title{{\small Phys. Rev. E, Vol. 53, 5382 (1996)}\\[0.5cm]
A model for collisions in granular gases}
\author{Nikolai V. Brilliantov\cite{bylinekolja},}
\affiliation{Moscow State University, Physics Department, Moscow 119899,
Russia}
\author{Frank Spahn, Jan--Martin Hertzsch,}
\affiliation{University Potsdam, Am Neuen Palais, D--14415 Potsdam,
Germany}
\author{Thorsten P\"oschel\cite{bylinetp}}
\affiliation{Humboldt--Universit\"at zu Berlin, Institut f\"ur Physik, 
Invalidenstra\ss e 110, \\ D--10115 Berlin, Germany}
\date{8 September 1994}
\begin{abstract}
We propose a model for collisions between particles of a granular
material and calculate the restitution coefficients for the normal and
tangential motion as functions of the impact velocity from
considerations of dissipative viscoelastic collisions. Existing models
of impact with dissipation as well as the classical Hertz impact
theory are included in the present model as special cases. We find
that the type of collision (smooth, reflecting or sticky) is
determined by the impact velocity and by the surface properties of the
colliding grains. We observe a rather nontrivial dependence of the
tangential restitution coefficient on the impact velocity.
\end{abstract}

\pacs{PACS numbers: 83.70.Fn, 62.40.+i, 81.40.Lm}
\maketitle
\section{Introduction}
A rich variety of systems one encounters in nature may be considered
as ``granular gas''~\cite{Campbell}. The most important difference
between a ``gas'' of granular particles and a regular gas is the
inelastic nature of the inter--particle collisions.  The steady
removal of kinetic energy from the granular gas due to dissipative
collisions causes a variety of non--equilibrium processes, which have
been subjects of experimental
(e.g.~\cite{exp1,exp2,exp3,exp4,exp5,exp6,exp7,exp8,exp9}) and
theoretical (e.g.~\cite{theo1,theo2,theo3,theo4,theo5})
interest. Particularly in recent time many of the experimental results
have been reproduced and investigated using various techniques such as
Cellular Automata (e.g.~\cite{CA1,CA2,CA3}), Monte--Carlo
Methods~\cite{MonteCarlo}, Lattice--Gas models~\cite{LatticeGas},
Molecular Dynamics in two~\cite{twoDMD1,twoDMD2,twoDMD3,twoDMD4} and
three~\cite{threeDMD1,threeDMD2,threeDMD3} dimensions and hybrid
methods~\cite{hybrid1,hybrid2,hybrid3,hybrid4}.

The loss of kinetic energy of a pair of inelastically colliding grains
can be described using the restitution coefficients for the normal and
tangential components of the relative motion $\epsilon^N$ and
$\epsilon^T$
\begin{mathletters}
\begin{eqnarray}
\left(\vec{g}^{\,N}\right)^\prime &=& - \epsilon^N~\vec{g}^{\,N} \qquad
              \left( 0 \le \epsilon^N \le 1 \right) \\
\left(\vec{g}^{\,T}\right)^\prime &=& \epsilon^T~\vec{g}^{\,T} \qquad
              \left( - 1 \le \epsilon^T \le 1 \right),
\end{eqnarray}
\label{restitutiondef}
\end{mathletters}
where $\vec{g}^{\,N}$ and $\vec{g}^{\,T}$ are the relative velocities
of the particles in normal and tangential direction before the
collision and $\left(\vec{g}^{\,N}\right)^\prime$,
$\left(\vec{g}^{\,T}\right)^\prime$ after the collision.

Recently, the collision properties of small spheres have been
investigated experimentally~\cite{Foerster}. These investigations have
shown that the type of the collision (sliding or sticking) depends on
the ratio of $g^N$ and $g^T$. The results were explained with
different models for each type, and the coefficients in these models
were independent from the velocity. On the other hand, laboratory
experiments with ice balls~\cite{Bridges} as well as with spheres of
other materials (for an overview see~\cite{impact}) have shown that
the normal restitution coefficient $\epsilon^N$ depends significantly
on the impact velocity.  As already seen, the tangential restitution
coefficient depends on the impact parameters as well. 

The behaviour of the sheared granular material may be significantly
different if the restitution coefficients depend on the impact
velocity. This dependence should be taken into account in order to
get an adequate model of the stress distribution~\cite{LunSavage}.
It is also known that the the parameters $\epsilon^N$ and $\epsilon^T$
crucially influence the global dynamics of granular systems
(e.g.~\cite{ArakiTremaine,WisdomTremaine}).

In the present study we investigate how the restitution coefficients
depend on the relative impact velocity. For the normal component of
the relative motion we derive an expression for the normal force
acting between the colliding particles, which accounts for the
dissipation in the bulk of material. One particular application of the
results presented here is the explanation of experiments with ice
balls~\cite{Bridges} which are of importance for the investigation of
the dynamics of planetary rings~\cite{Planetenbuch}.  A static model
for the inelastic impact of metal bodies was presented
in~\cite{Johnson} which based on the assumption of fully plastic
indentation and constant mean contact pressure and led analytically to
a proportionality $\epsilon^N \propto \left( g^N \right)^{-1/4}$ for
arbitrary material constants.  In contrary, our quasistatic approach
does not request other additional assumptions and can be adapted to
different experimental results by changing the coefficients in the
differential equation which describes the time dependence of the
deformation. From these coefficients, material coefficients can be
estimated~\cite{Caputh}.

Our result contains the Hertz theory of elastic impact~\cite{Hertz}
and the theory of the viscoelastic impact by Pao~\cite{pao} as special
cases.  For the tangential component of the relative motion we
consider a microscopic model of the contact of colliding particles.
We derive a mean--field expression for the tangential inter--particle
force. The result contains the model of the tangential force of
colliding particles by Haff and Werner~\cite{CundallStrack,HaffWerner}
as a special case, and we are able to treat different tangential
collisional behaviors within the framework of one single model.

In section~\ref{CollisionModelSec} we formulate the collision model
and derive the equations for the normal and tangential components of
the relative motion of the colliding grains. In
section~\ref{ResultsDiscussionSec} we present the results for the
restitution coefficients for the proposed model and discuss the
dependence of the coefficients on the components of the impact
velocity. A model for the dynamics of granular materials is proposed.
In the last section~\ref{conclusion} we summarize the results.
Details of the derivations are given in the appendixes~\ref{AppendixA}
and~\ref{AppendixB}.

\section{The collision model}
\label{CollisionModelSec}
We consider the inelastic collision between two spherical particles
$i$ and $j$. The values $\vec{r}_i$, $R_i$, $\dot{\vec r}_i$,
$\dot{\vec{\omega}}_i$, $m_i$, and $J_i$ are the position of the
center of sphere $i$, its radius, velocity, angular velocity, mass and
momentum of inertia, respectively. The relative velocity of the
surfaces of the colliding particles at the point of contact is given
by (e.g.~\cite{CundallStrack,impact})
\begin{eqnarray}
\vec{g}_{ij} & = & \left(\dot{\vec{r}}_i - \vec\omega_i \times 
                R_i~\vec{n} \right) - \left(\dot{\vec{r}}_j + 
                \vec\omega_j \times R_j~\vec{n}\right)
                \nonumber\\
            & = & \dot{\vec{r}}_i - \dot{\vec{r}}_j - 
                R_i~\vec{\omega}_i \times \vec{n} - 
                R_j~\vec{\omega}_j \times \vec{n}~,
\label{gijpunkt}
\end{eqnarray}
with $\vec{n} = \frac{\vec{r}_i - \vec{r}_j}{\mid \vec{r}_i -
\vec{r}_j \mid}$. Introducing the dimensionless moment of inertia
$\hat J_i$, the effective mass $m_{ij}^{\mbox{\it\footnotesize eff}}$
and the effective radius $R_{ij}^{\mbox{\it\footnotesize eff}}$
\begin{mathletters}
\begin{eqnarray}
\hat{J}_i &=& \frac{J_i}{m_i~R_i^2}\\
m_{ij}^{\mbox{\it\footnotesize eff}} &=& \frac{m_i m_j}{m_i+m_j}\\
R_{ij}^{\mbox{\it \footnotesize eff}} &=& \frac{R_i R_j}{R_i+R_j}
\end{eqnarray}
\end{mathletters}
one obtains Newtons equations for the translational and rotational
motion
\begin{equation}
\frac {\mbox{d}\vec{g}_{ij}} {\mbox{d}t} = 
   \frac{\vec{F}_{ij}}{m_{ij}^{\mbox{\it \footnotesize eff}}} +
   \left(\frac{1}{\hat{J}_i m_i} + \frac{1}{\hat{J}_j m_j}\right) 
   \left( \vec{n} \times \vec{F}_{ij} \right) \times \vec{n}~.
\end{equation}
The force $\vec{F}_{ij}$ acting between the particles during collision
consists of the normal component $\vec{F}_{ij}^N =
\vec{n}\left(\vec{n} \cdot \vec{F}_{ij}\right)$ and the tangential
component$\vec{F}_{ij}^{\,T}=\vec{F}_{ij}-\vec{F}_{ij}^{\,N}$.
Introducing the corresponding components $\vec{g}_{ij}^{\,N}$ and
$\vec{g}_{ij}^{\,T}$ of the relative velocity $\vec{g}_{ij}$ and with
\begin{equation}
\kappa_{ij}^{-1}=1+\frac{m_i \hat{J}_i+m_j\hat{J}_j}
                {\hat{J}_i \hat{J}_j \left( m_i+m_j \right)}
\label{kappadef}
\end{equation}
we rewrite eq.~(\ref{gijpunkt}) omitting the indexes $ij$:
\begin{mathletters}
\begin{eqnarray}
\dot{\vec{g}}^{\,N} &=& \vec{F}^N / m^{\mbox{\it \footnotesize eff}}
   \label{gna}\\
\dot{\vec{g}}^{\,T} &=& \frac{1}{m^{\mbox{\it \footnotesize eff}} 
                      \kappa}\,\vec{F}^T \label{gnb}~.
\end{eqnarray}
\end{mathletters}
Using eqs.~(\ref{restitutiondef}) the energy loss during the collision
is
\begin{equation}
\Delta Q = \frac{m^{\mbox{\it \footnotesize eff}}}{2}
           \left( \vec{g}^{\,N}\right)^2 \left( \left( 
            \epsilon^N\right)^2-1\right) +
           \frac{m^{\mbox{\it \footnotesize eff}}}{2}
           \kappa~\left( \vec{g}^{\,T}\right)^2\left(\left(
           \epsilon^T\right)^2-1\right)~.
\end{equation}
The energy is conserved during the collision if $\epsilon^N=1$ and
$\epsilon^T=\pm 1$. In these cases there is a completely elastic
rebound for the normal component and either completely elastic rebound
(rough spheres) or frictionless slipping (smooth spheres) for the
tangential component.

\subsection{Normal motion}
We assume that the colliding particles begin to touch each other at
the time $t=0$ with the relative normal velocity $\vec{g}^{\,N}$. When
we introduce the deformation (or ``compression'')
 
\begin{equation}
\xi(t)=R_i+R_j -\left(\mid\vec{r}_{i}(t)-\vec{r}_{j}(t)\mid\right)
\end{equation}
this velocity can be written as
$g^N= \mid\vec{g}^{\,N} \mid = \dot \xi$. 

Thus from
eq.~(\ref{gna}) we obtain the equations
\begin{eqnarray}
\label{gn1}
\ddot{\xi}(t) &=& F^N \left(\xi(t)\right)/
                m^{\mbox{\it\footnotesize eff}}\\
\dot{\xi}(0) &=& g^N\nonumber \\
\xi(0) &=& 0\nonumber~.
\end{eqnarray}
The normal force $F^N$ consists of an elastic, conservative part due
to the deformation $\xi$ of the particles and a viscous part due to
dissipation of energy in the bulk of the particle material, depending
on the deformation rate $\dot \xi$. For the conservative part 
Hertz's theory of elastic contact~\cite{Hertz} gives for spherical
particles
\begin{equation}
F^N_{(el)}(\xi) = \frac{2~Y}{3 ~\left( 1-\nu ^2\right)}
                  \sqrt{R^{\,\mbox{\it \footnotesize eff}}}
                  ~\xi^{3/2}~,
\label{hertz}
\end{equation}
where $Y$ and $\nu $ are the Young modulus and the Poisson ratio for
the material the particles consist of. This relation between the
elastic component of the force and the deformation is valid for the
quasistatic regime of the collision, i.e. when inertial and relaxation
effects may be neglected (see appendix~\ref{AppendixB}).

The existing phenomenological expressions for the dissipative part of
the normal force which are either linear in the deformation rate $\dot \xi$
(e.g.~\cite{CundallStrack,Dilley}) or quadratic~\cite{Poeschl},
however, do not agree satisfactorily with the experimental data for the
normal restitution coefficient~\cite{Bridges}. Pao~\cite{pao} extended
the Hertz theory of impact for the viscoelastic case, where, however,
the dependence of the bulk dissipation on the dilatation rate was
neglected.  In this theory memory effects in the dissipative processes
were taken into account.  Although the latter approach is not
self--consistent (see appendix~\ref{AppendixB}), it predicts a
power--law dependence of the dissipative force on the deformation
rate, yielding an exponent similar to that one for the quasistatic
collision.  In the present study we develop a self--consistent
quasistatic approximation to calculate the normal force acting between
colliding viscoelastic particles. The quasistatic approximation is
valid when the characteristic relative velocity of the granular
particles is much less than the speed of sound in the material which
is satisfied for many experimental situations even in astrophysical
systems such as planetary rings~\cite{SpeedPlanetaryRings}. For the
duration of the collision it is necessary to exceed significantly the
viscous relaxation time in the material of colliding particles (see
appendix~\ref{AppendixB}).

Different from the approaches~\cite{pao,impact} we take into account
both components of the dissipative force, arising from the shear
strain rate as well as from the dilatation rate, which are both of
comparable importance for the normal component of the relative motion.
From the equation of motion for the viscoelastic continuum we find
the general relation between the dissipative part of the normal force
and the deformation rate. We show that memory effects in dissipative
processes may be neglected in the case of a self--consistent
quasistatic approximation. Since the calculation of the dissipative
part of the normal force is rather straightforward, we present only
the main idea of the derivation and refer to appendix~\ref{AppendixA}
for further details.  In appendix B the conditions for the validity of
the quasistatic approach are given.

The total normal force acting between viscoelastic particles may be
derived from a stress tensor combined of an elastic and a dissipative
part~\cite{Landau}
\begin{equation}
\hat{\sigma}=\hat{\sigma}_{(el)}+\hat{\sigma}_{(dis)}
\end{equation}
with
\begin{mathletters}
\begin{eqnarray}
\hat{\sigma}_{(el)} &=& E_1~ \left[\frac{1}{2}
                    \left\{\nabla\circ\vec{u} + 
                    \vec{u}\circ\nabla\right\} -
                    \frac{1}{3}\hat{I} \,\nabla\cdot\vec{u}\right] + 
                    \nonumber \\
&& E_2\,\hat{I}\, \nabla \cdot \vec{u} \\
\hat{\sigma}_{(dis)} &=& \eta_1~ \left[\frac{1}{2}
                    \left\{\nabla\circ\dot{\vec{u}} + 
                    \dot{\vec{u}}\circ\nabla\right\} -
                    \frac{1}{3}\hat{I}\,\nabla
                    \cdot \dot{\vec{u}} \right] +
                    \nonumber \\
&& \eta_2\,\hat{I}\, \nabla \cdot \dot{\vec{u}}~.
\end{eqnarray}
\label{stresstensor}
\end{mathletters}
The displacements in the material is denoted by $\vec{u}$ and
$\hat{I}$ is the unit tensor. $E_{1/2}$ and $\eta_{1/2}$ are the
elastic and the viscous constants of the particle material
\begin{mathletters}
\begin{eqnarray}
E_1 &=& \frac{Y}{1+\nu }\\
E_2 &=& \frac{Y}{3\left(1-2\nu \right)}~.
\end{eqnarray}
\end{mathletters}
In the quasistatic regime (appendix~\ref{AppendixB}) the displacement
field $\vec{u} \left(\vec{r}, t\right)$ can be approximated by that of the static
problem $ \vec{u} \left(\vec{r}\right)$. It is completely determined by the
elastic component of the inter--particle force (\ref {hertz}).  Thus,
the displacement velocities can be written as
\begin{equation}
\dot{\vec{u}}\left(\vec{r},t\right)\simeq\dot{\xi}\,
\frac{\partial{}}{\partial{\xi}}
                            \,\vec{u}_{(el)}\left( \vec{r}, \xi \right)~,
\label{displvelocities}
\end{equation}
where $\vec{u}_{(el)}\left(\vec{r}, \xi \right)$ is the solution of the static
(elastic) contact problem. This expression depends parametrically on
the deformation $\xi$ and the dissipative part of the stress tensor
becomes
\begin {eqnarray}
\hat{\sigma}_{(dis)} &=& 
      \dot {\xi} \,\frac{\partial {}}{\partial {\xi}}\,
      \left\{ \eta_1\,\left[\frac{1}{2}
      \left(\nabla\circ\vec{u}_{(el)}  + 
      \vec{u}_{(el)}\circ\nabla\right) -\nonumber\right.\right.\nonumber\\
&& \left.\left. \frac{1}{3}\hat{I}\,\nabla\cdot \vec{u}_{(el)}\right] + 
     \eta_2\,\hat{I }\,\nabla \cdot \vec{u}_{(el)}\right \}~.
\end{eqnarray}
The calculations can 
be significantly simplified when we notice that the
elastic and the dissipative parts of the stress tensor are related in
the quasistatic regime (see eqs.~(\ref{stresstensor}) and
(\ref{displvelocities}))
\begin{equation}
\sigma_{(dis)}= \dot {\xi} \,\frac{\partial {}}{\partial {\xi}} 
                 \hspace{0.1cm} \sigma_{(el)}
                 ~~~~~ ,~~~~ \left(E_1\leftrightarrow \eta_1,\, 
                 E_2 \leftrightarrow \eta_2\right)~.
\end{equation}
Therefore the impact problem for the viscoelastic particles in the
quasistatic regime can be mapped onto the corresponding problem for
elastic particles. Performing calculations similar to that of the
elastic case (for details see~\cite{Hertzsch} and
appendix~\ref{AppendixA}) one can find an expression for the
dissipative part of the normal force:
\begin{eqnarray}
F^N_{(dis)} &=& \frac{Y}{ \left( 1-\nu ^2\right)}
            \sqrt{R^{\,\mbox{\it\footnotesize eff}}}~A \sqrt{\xi}
            \, \dot {\xi}\label{forcedis}\\
A &=& \frac{1}{3} \, \frac{ \left(3 \eta_2-\eta_1\right)^2}
      {\left(3\eta_2+2\eta_1\right)}\, 
      \left[\frac{\left (1-{\nu}^2 \right )\left (1-2 \nu \right)}
      {Y\,\nu^2}\right]\nonumber~.
\end{eqnarray}
From eqs.~(\ref{forcedis}) and (\ref{hertz}) we obtain for the normal
component of the relative motion
\begin{equation}
\ddot \xi +\frac{2~ Y\sqrt{R^{\,\mbox{\it \footnotesize eff}}}}{
3~ m^{\mbox{\it \footnotesize eff}}\left( 1-\nu ^2\right) }\left(
 \xi^{3/2} +\frac32\,\, A\, \sqrt{\xi}\, \dot {\xi}\right)=0
\label{xitaylor}
\end{equation}
with the initial conditions $\dot{\xi} \left( 0 \right) = g^N$, $\xi
\left( 0 \right) = 0$. 
In the case of $A \dot{\xi} \ll \xi~$
equation~(\ref{xitaylor}) results from a Taylor--expansion of
\begin{equation}
\ddot \xi +\frac{2~ Y\sqrt{R^{\,\mbox{\it \footnotesize eff}}}}
     {3~ m^{\mbox{\it \footnotesize eff}}\left( 1-\nu ^2\right)}
     \left(\xi +\, A\, \dot {\xi}\right )^{3/2}=0
\label{xifull}
\end{equation}
which formally coincides with the corresponding equation for the
elastic problem, provided that $\xi$ is substituted by $\xi+A\,
\dot{\xi}$. 

It has to be noted, that $\xi$ has its minimum at the beginning of the
collision where $\dot \xi$ takes its maximum. Hence, the condition $A
\dot{\xi} \ll \xi $ is not provided at the very beginning of the
contact. On the other hand, the good confirmation of
experimental facts~(\cite{Bridges}) by the numerical solution of
equation (\ref{xifull}) points to its suitability for at least the
rest of the collision time span.

Taking into account $\left( \dot{g}^N \right)^{\prime }={\dot \xi (}t_c)$ 
($t_c$ -- the duration of the collision), the normal
restitution coefficient is obtained from 
\begin{equation}
\epsilon^N={\dot \xi (}t_c)/{\dot \xi (0}).
\label{epsn}
\end{equation}

\subsection{Tangential motion}
In the idealized model the surface of contact between the spheres $S$
is a perfectly flat circular area with radius $R_S=\sqrt{2 R^{\,\mbox{\it
\footnotesize eff}}\, \xi (t)}$. For the description of the tangential forces
between the surfaces we follow a current model of tribology
(e.g.~\cite{tribology1,tribology2}) where the apparent surface of contact is built
up of a large number of hierarchically ordered asperities
varying in shape and size by
several decades. For the processes of the momentum
transmission we will take into account only the largest--scale
asperities (``primary asperities''). The surface asperities do not
affect the normal motion, if they are small enough 
(see appendix~\ref{AppendixA}), however they are responsible 
for the tangential forces, acting between the colliders. Here 
we consider  a simplified mean-field approach and introduce the normal 
$\overline{\sigma^N}$ and tangential stress $\overline{\sigma^T}$
averaged over the contact area. Further we define the normal component
of the total contact area of the asperities of both spheres $S^N$
which is responsible for the transmission of the normal force.
Correspondingly the tangential projection of the area $S^T$ is
responsible for the transmission of the tangential force.
These surfaces are related to the apparent contact area
by the relations~\cite{Greenwood}
\begin{mathletters}
\begin{eqnarray}
S^N(t)&=&f^N\left(\overline{\sigma^N}\right)~ S(t) \\
S^T(t)&=&f^T\left(\overline{\sigma^N}\right)~ S(t)~ ,
\label{SNST}
\end{eqnarray}
\end{mathletters}
where the coefficients $f^N$ and $f^T$ depend on the average normal
stress $\overline{\sigma^N}$. When the spheres begin to touch each
other, i.e. $S=0$ and $\overline{\sigma^N}=0$, we find $f^N(0)=0$ and
$f^T(0)=0$. We expand the coefficients in eq.~(\ref{SNST}) with
respect to $\overline{\sigma^N}=0$. The linear expansion yields
for the tangential component of the surface
\begin{eqnarray}
S^T(t) &=& \phi^T~\overline{\sigma^N}~S(t)\\
\phi^T &=& \left[\frac{\partial f^T}{\partial \overline{\sigma^N}}
         \right]_{\overline{\sigma^N}=0}\nonumber~.
\end{eqnarray}
For a given model of the sizes and shapes of the asperities one can
calculate the value of $\phi^T$ ~\cite{Greenwood}. In the case that the
heights of the asperities obey a Gaussian probability distribution
with mean value $L$ one finds
\begin{equation}
\phi^T \propto \sqrt{L}~.
\label{phitproptosqrtl}
\end{equation}
For the average size of the asperities $L$ of the surfaces
the mean field approach yields the average shear deformation
$\overline{\eta}=b~\frac {\zeta}{L}$. The values $\zeta$ and
$b$ are the relative tangential shift of the particle-surfaces and
a form-factor, respectively.
We assume that the stress is uniformly
distributed over the entire surface and find
\begin{equation}
\overline{\sigma^T} = \frac{Y}{1+\nu}\overline{\eta}
                    = \frac{Y~\zeta}{\left(1+\nu\right)~L}~b~.
\end{equation}

The linear relation between $\overline{\sigma^T}$ and $\overline{\eta
}$ holds only for the elastic regime, i.e. only if
$\overline{\sigma^T}$ does not exceed some critical value
$\sigma^T_*$ which is a specific material constant. If the shear
stress exceeds this threshold $\sigma^T_*$, the asperity which hinders
the tangential relative motion of the surfaces, is assumed to break
resulting in a sudden release of the shear stress.  At the same time
the surfaces are shifted macroscopically with respect to each other by
\begin{equation}
\zeta_0=\frac{L~\left(1+\nu\right)} {b~Y}~\sigma_*^T~,
\label{zeta0}
\end{equation}
and one finds
\begin{eqnarray}
\overline{\eta }\left(\zeta \right) &=& \eta_*\left(\frac{\zeta}{\zeta_0}
          -\left\lfloor \frac{\zeta}{\zeta_0}\right\rfloor \right)
          \label{etaquer}\\
\eta_* &=&\frac{b\zeta_0}{L} \nonumber
\end{eqnarray}
where $\lfloor x\rfloor$ denotes the integer of $x$.  The breaking of
the asperities dissipates the energy which was previously stored in
the elastic stress, i.e. fracturing of the asperities is the
elementary dissipative process in the tangential motion. From
eq.~(\ref{etaquer}) we obtain the shear stress as a function of
the tangential displacement
\begin{equation}
\overline{\sigma^T}\left(\zeta\right)=\sigma^T_*\left(\frac{\zeta}{\zeta_0} -
           \left\lfloor\frac{\zeta}{\zeta_0}\right\rfloor
\right)
\end{equation}
and the tangential component of the inter--particle force
\begin{eqnarray}
F^T &=& -S^T\overline{\sigma^T}\left(\zeta\right)
     = -\phi^T\overline{\sigma^N}S\sigma^T_*
       \left(\frac{\zeta}{\zeta_0} -
       \left\lfloor \frac{\zeta}{\zeta_0}\right\rfloor \right)
       \nonumber \\
    &=& -\mu F^N\left(\frac{\zeta}{\zeta_0} -
       \left\lfloor\frac{\zeta}{\zeta_0}\right\rfloor \right)~.
\label{FT}
\end{eqnarray}
where $F^N=\overline{\sigma^N}S$ is the normal component of the
inter--particle force and $\mu =\phi^T\sigma^T_*$.

It may be shown, that a more refined mean-field approach, which does 
not use the assumption of the uniformly distributed stress over the 
contact area, leads to the same eq.~(\ref{FT}) for the tangential 
motion.

From eq.~(\ref{FT}) follows the condition for the maximum tangential
force
\begin{equation}
F^T_{\mbox{\it \footnotesize max}}=\mu F^N~.
\end{equation}
Thus our model reproduces the Coulomb friction law~\cite{Coulomb} with
the friction coefficient $\mu$, expressed in terms of microscopic
parameters.The model for the tangential motion is very similar to the
extensively investigated one dimensional model by Burridge and
Knopoff~\cite{BurridgeKnopoff,CalsonLanger} intended to model earthquakes.

With $g^T(t)=\dot \zeta (t)$, eq.~(\ref{kappadef}) and
$F^N=-m^{\mbox{\it \footnotesize eff}} \ddot \xi (t)$ the tangential
motion is governed by the differential equation
\begin{equation}
\ddot \zeta - \frac{\mu}{\kappa} \ddot \xi \left( t \right)
    \left( {\zeta\over \zeta_0} -
    \left\lfloor {\zeta\over\zeta_0} \right\rfloor \right) = 0
\label{zeta}
\end{equation}
with the initial conditions $\dot\zeta\left(0\right)=g^T$ and
$\zeta\left(0\right)=0$. The value of $\ddot \xi (t)$ is given by
eqs.~(\ref{xitaylor}) and (\ref{xifull}). Then the tangential
restitution coefficient reads
\begin{equation}
\epsilon^T={\dot\zeta }\left(t_c\right)/{\dot \zeta }(0)~.
\label{epst}
\end{equation}

\section{Results and discussion}
\label{ResultsDiscussionSec}
\begin{figure}[htbp]
\centerline{ \psfig{figure=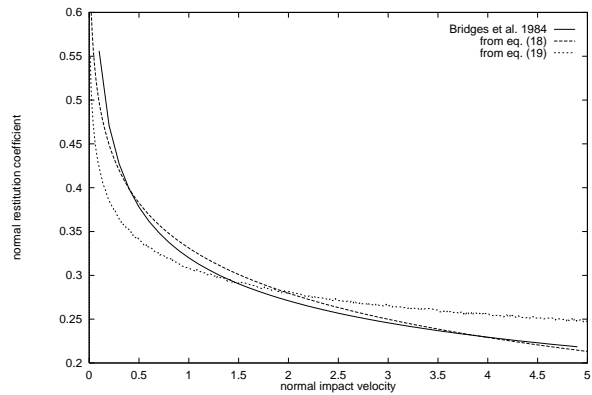,width=8cm,angle=270}}
\caption{The normal restitution coefficient $\epsilon^N$
versus the normal component of the impact velocity $g^N$ measured in
${cm }\ {s}^{-1}$ according to eqs.~(18)
 and (19).
The dashed line denotes the dependence
$\epsilon^N(g^N)$ measured by Bridges et al.~[33]}
\label{fig=enorm}
\end{figure}

The obtained equations for the normal (eq.~(\ref{xifull}),
(\ref{xitaylor})) and tangential motion (eq.~(\ref{zeta})) have been
solved numerically using a Runge--Kutta Method of fourth order with
adaptive step size~\cite{recipes}. The restitution coefficients
$\epsilon^T$ and $\epsilon^N$ have been calculated as functions of the
normal and tangential relative velocities $g^T$, $g^N$. For the
integration we used the parameters of ice at low
temperatures~\cite{LandoltBoernstein}: Young modulus $Y= 10~GPa$, the
Poisson ratio $\nu = 0.3$, the particle size $R= 10^{-2}~m$ with
density $\rho=10^3~kg~m^{-3}$. The coefficient $A$ in
eq.~(\ref{xitaylor}) was considered to be a fit parameter, due to lack
of information about the dissipative coefficients $\eta_1$ and
$\eta_2$. Fig.~\ref{fig=enorm} shows the numerical result of our model
for the normal restitution coefficient $\epsilon^N$ as a function of
the normal relative velocity $g^N$ compared to experimental data for
the collision of spherical ice particles with an ice
wall~\cite{Bridges}. The experimental results are well
reproduced by our model.

\begin{figure}[htb]
\centerline{\psfig{figure=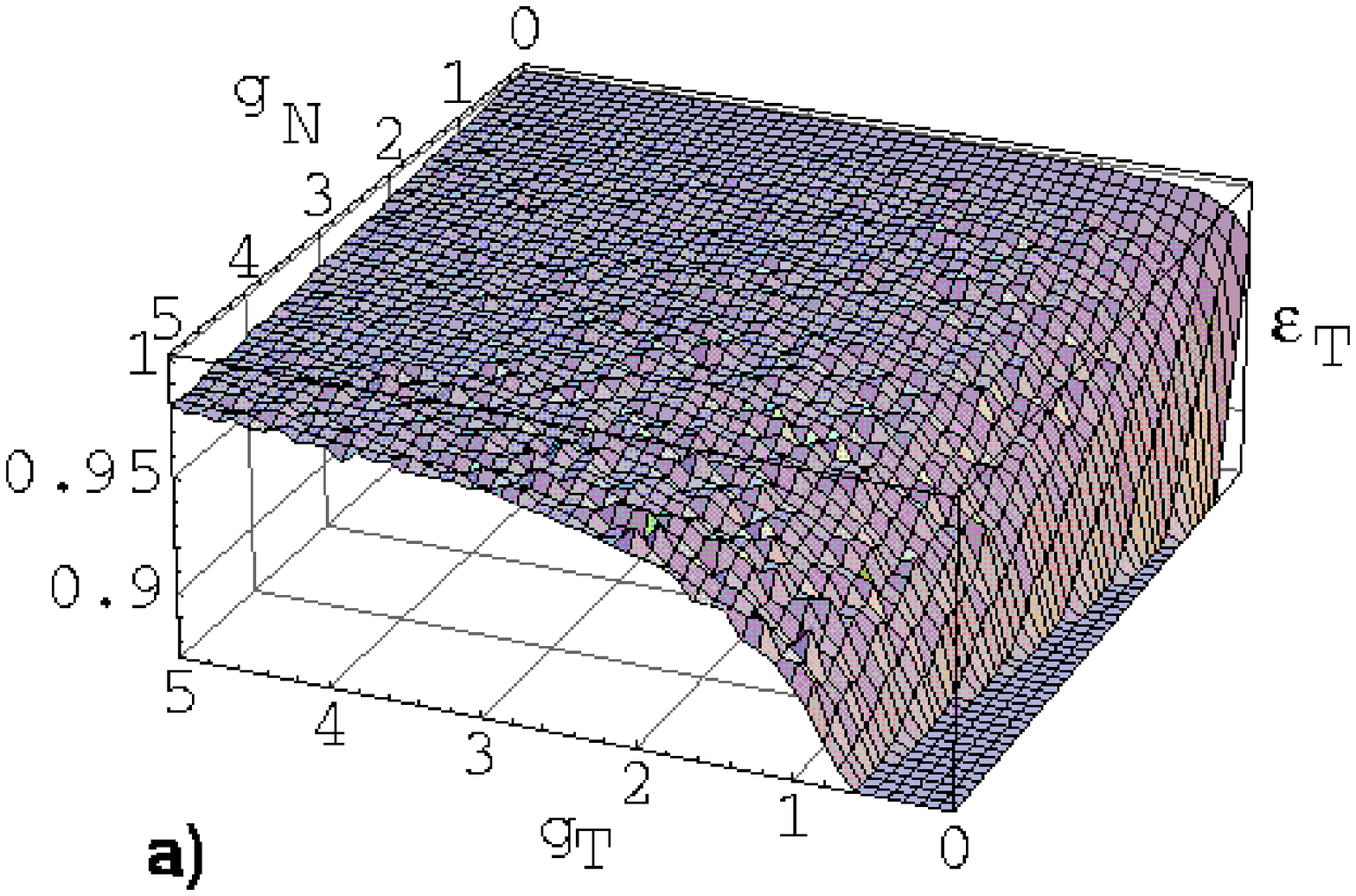,width=6.7cm}}
\centerline{\psfig{figure=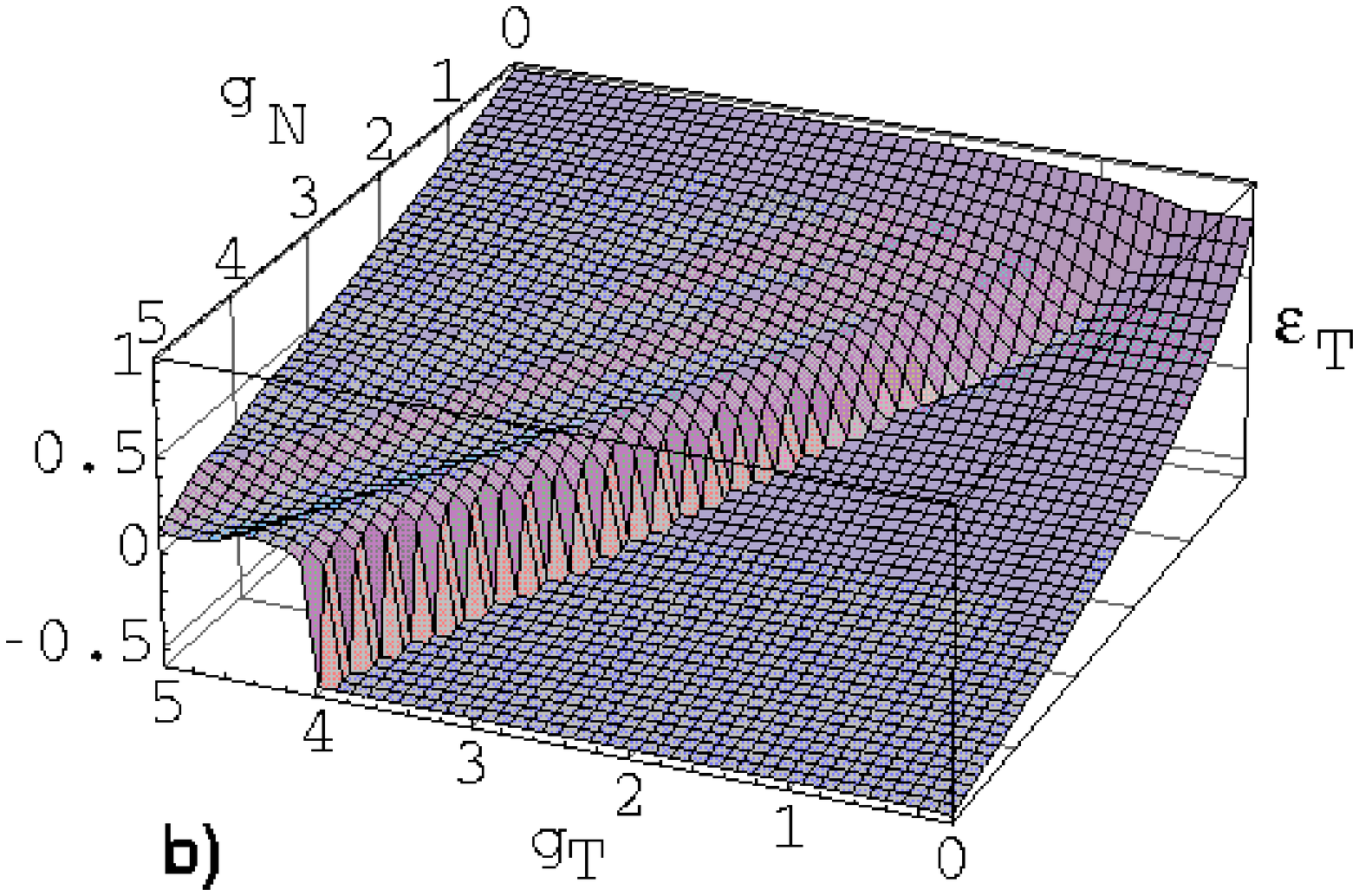,width=6.7cm}}
\centerline{\psfig{figure=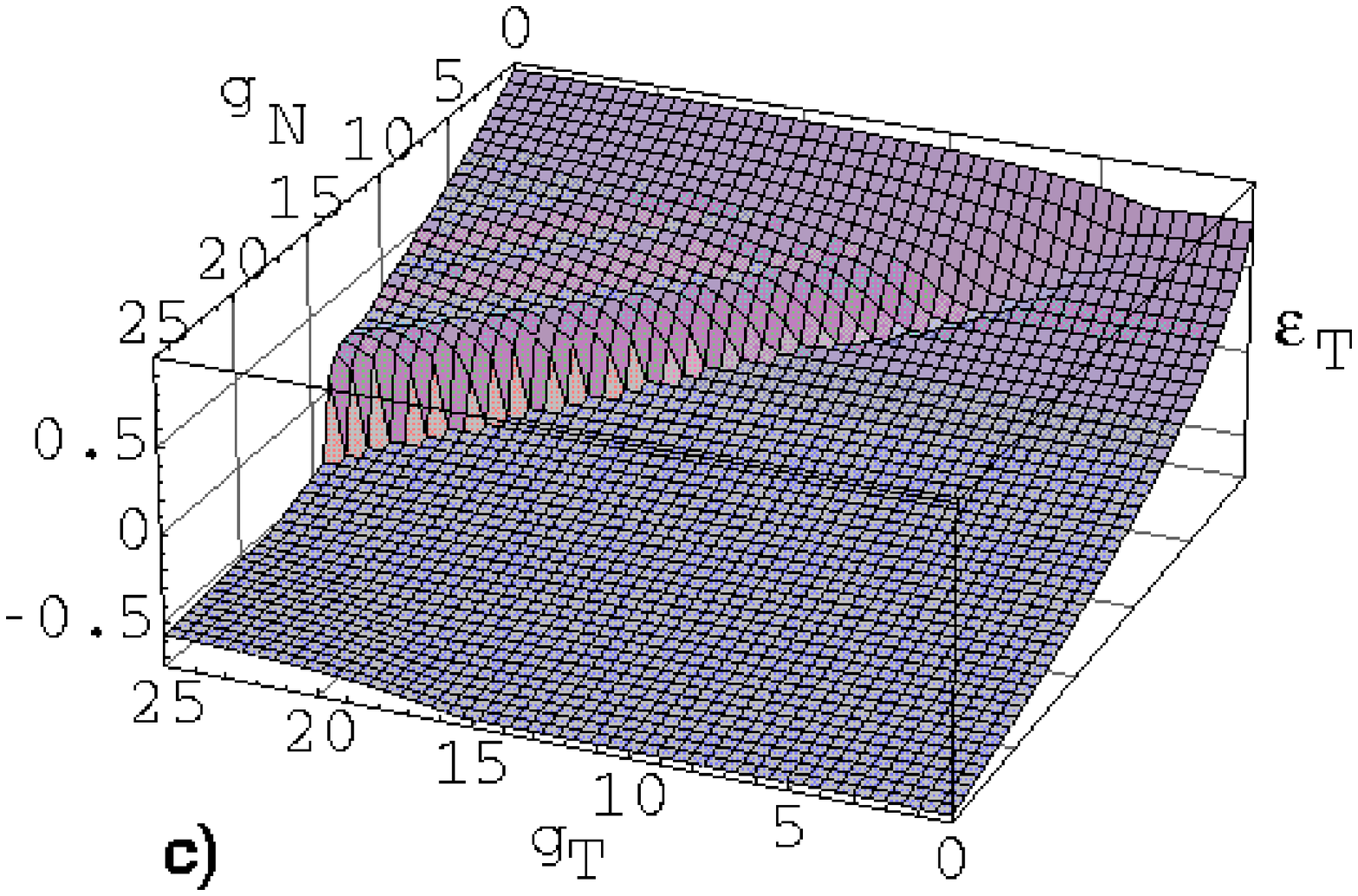,width=6.7cm}}
\caption{The stereographic projection of the tangential
  restitution coefficient $\epsilon^T$ versus the plane $g^N$--$g^T$
  of the tangential and normal components of the impact velocity.  The
  three parts of the figure belong to different values of the ``size''
  of the asperities: (a) $\zeta _0=10^{-7}~R^{\mbox{\it \footnotesize
      eff}}$, (b) $\zeta _0=2\cdot 10^{-4}~R^{\mbox{\it \footnotesize
      eff}}$, (c) $\zeta _0=10^{-3}~R^{\mbox{\it \footnotesize eff}}$.}
\label{fig=etansurface}
\end{figure}

For the investigation of the tangential restitution coefficient of
colliding homogeneous spheres ($J=\frac{2}{5}mR^2$,
$\kappa=\frac{2}{7}$) we have chosen the Coulomb friction coefficient
from the interval $\mu\in [10^{-2}\dots 1]$. The value of $\sigma^T_*$
is a material constant. With eq.~(\ref{phitproptosqrtl}) and the
definitions of $\xi_0$ and $\eta_*$ we estimate $\zeta_0$ which
characterizes the size of the surface asperities via
$\mu=\alpha\,\,\sqrt{\zeta_0}$. In the numerical calculation we have chosen
$\alpha=1$. The results are shown in fig.~\ref{fig=etansurface}. The
tangential restitution coefficient $\epsilon^T$ is drawn versus the
plane defined by the tangential and normal velocities $g^T$ and $g^N$. 
The three plots correspond to
the values of the asperity sizes $\zeta_0=\left(10^{-7}; 2\cdot
10^{-4};~10^{-3}\right)~ R^{\,\mbox{\it \footnotesize eff}}$,
respectively.

The obvious common feature of all cases is sliding of the surfaces 
($\epsilon^T>0$) for small $g^N$ and large $g^T$.
This is quite plausible since smaller impact velocity $g^N$ corresponds
to a smaller normal acceleration and thus, to a smaller value of the 
maximal tangential force,  eq.~(\ref{FT}). As a result 
$\epsilon^T \rightarrow 1$ at  $g^N  \rightarrow 0$ due to vanishing 
tangential acceleration. At the same time, for the high tangential 
velocity, $g^T(0) \gg 1$, ($g^N(0) \simeq 1$) sliding occurs owing to 
a considerable breaking of the asperities. 
The area of the sliding phase in the
$g^N$--$g^T$ plane depends on the size $\zeta_0$ of the asperities.

In the case of $\zeta_0=10^{-7}~R^{\mbox{\it \footnotesize eff}}$
sliding occurs in the entire velocity range according to values $0.85
\le \epsilon^T \le 1$. The small asperities are not able to cause a
sufficient torque to change considerably the spin of the
individual particles. Here we are close to the case of ideal smooth
spheres where no change of the tangential motion is expected
($\epsilon^T=1$).

In the other two cases $\zeta_0=\left(2\cdot 10^{-4};~10^{-3}\right)~
R^{\,\mbox{\it \footnotesize eff}}$ one recognizes two phases:
\begin{enumerate}
\item Sliding $\epsilon^T>0$ at small $g^N$ and high $g^T$
\label{sliding}
\item Reversal of the spin of either particles $\epsilon^T<0$ at small
$g^T$ and higher $g^N$.
\label{reversal}
\end{enumerate}
Case~\ref{sliding} corresponds to the effect discussed in the context
of $\zeta_0=10^{-7}$. Despite of being far from rather smooth spheres,
the small tangential force originated from small
$g^N$, changes the velocity $g^T$ only slightly. Hence one has
$\epsilon^T > 0$, which is also the case for high velocities $g^T$
where the asperities break. In the case~\ref{reversal} we have the
other extreme: a high normal acceleration causes a tangential force
which is high enough to change the sign of
$g^T$ as long as the asperities do not break (small $g^T$). A complete
reversal of the tangential velocity according to
$g^T(0)\rightarrow~-g^T\left(t_c\right)$ is not possible because of the
dissipation arising of the bulk viscosity of the material which enters
the normal as well as the tangential forces (see
eqs.~(\ref{xitaylor}),(\ref{zeta})).

Both types of behavior of the tangential motion are separated by a
sharp transition at $g^T=g^T_{(cr)}$ where the asperities begin to break
(see surface plots for $\zeta_0=\left(2\cdot 10^{-4};~10^{-3}\right)~
R^{\,\mbox{\it \footnotesize eff}}$). The higher $g^N$ the larger the
critical tangential velocity $g^T_{(cr)}$. A higher normal velocity
$g^N$ causes a stronger counteracting force $F^T$ and thus a larger
tangential impact speed $g^T$ is necessary to reach the critical
deformation where the asperities break. Both cases ($\zeta_0=\left(2\cdot
10^{-4};~10^{-3}\right)~ R^{\,\mbox{\it \footnotesize eff}}$) reveal
similar qualitative behavior but the ranges of different type of
motion (\ref{sliding},~\ref{reversal}) cover different areas in the
$g^T$--$g^N$ plane.

The results show that our model includes a continuous transition from
the limit case of rough spheres ($\epsilon^T\rightarrow~-1$) to the
limit case of smooth spheres ($\epsilon^T\rightarrow~1$). In the
literature of the dynamics of granular material an alternative step
function is widely used for the tangential force~\cite{CundallStrack}
\begin{equation}
F^T=\min \left\{- \gamma_s m^{\mbox{\it \footnotesize eff}} \left| g^T
\right| ,~\mu F^N \right\}~.
\end{equation}

The numerical evaluation of the considered model (fig.~\ref{fig=etansurface})
reveals surprising behavior of the tangential restitution coefficient
$\epsilon^T$ as a function of the normal velocity $g^N$ at fixed
tangential velocity $g^T$. (This effect is noticeable for the largest
values of $\zeta_0$.) At low and moderate $g^N$, $\epsilon^T$ first
decreases with increasing $g^N$ down to its minimal negative value
in a manner discussed above, but when $g^N$ exceeds some threshold
(approximately of several $g^T$), it starts to increase up to zero at
very high values of $g^N$. This effect may be explained as follows:
For high values of $g^N$ the average normal force is large and 
causes thus a large tangential force, which can effectively decelerate the
initial tangential velocity without switching to the sliding
regime. 

Calculating the restitution coefficients $\epsilon^T$, $\epsilon^N$,
(in the limits of our model) we obtain a complete description of
binary collisions. Therefore one can determine the dynamics for
moderately dense granular gases, where an evolution occurs via a
sequence of binary collisions. For such systems we have the following
Boltzmann equation for the one--particle distribution function
\begin{widetext}
\begin{equation}
\label{boltz}
\left(\frac {\partial{}}{\partial t}+\vec{v}_1 \cdot \nabla
\right)f(1)  =   \int d \vec{v}_2 \int d\vec{\omega}_2\int d \vec{n}
 \left|\vec{g}^{\,N} \cdot \vec{n} \right| \Theta \left(\vec{g}^{\,N}
\cdot \vec{n}
\right) \left[\frac{f\left(1'\right)f\left(2'\right)}{\left(\epsilon^N~\epsilon^T\right)^2}-f\left(1\right)f\left(2\right) \right]
\end{equation}
\end{widetext}
with $\Theta(x)$ given by 
\begin{equation}                                                     
\Theta(x)=\left\{\begin{tabular}{lll}
1 & \mbox{~for~} & $x \ge 0$ \\       
0 & \mbox{~for~} & $x < 0 $~.   
\end{tabular} \right.
\label{Theta}                                   
\end{equation}
and with the common notations, e.g. $(1)= \left \{\vec{r}_1, \vec{v}_1,
\vec{\omega}_1 , t \right \}$. The velocity and angular velocity of the
first particle after the collision $\vec{v}_1 ^{\,\prime}$ and $\vec
{\omega}_1^{\prime}$ can be expressed in terms of the pre--collisional
values via the relations
\begin{widetext}
\begin{mathletters}
\begin{eqnarray}
\vec{v}_1 ^{\,\prime} &=& \vec{v}_1 + 
           \frac{m^{\mbox{\it \footnotesize eff}}}{2m_1}~
           \left\{\left[\epsilon^T \left(g^N, g^T\right)-1\right]
           \vec{g}^{\,T} -
           \left[\epsilon^N \left(g^N, g^T\right)+1\right]\vec{g}^{\,N} \right\} \\
\vec{\omega}_1^{\prime} &=& \vec{\omega}_1 + 
           \frac{m^{\mbox{\it\footnotesize eff}}}{2m_1} R~ 
           \vec{n} \times \left\{\left[\epsilon^T \left(g^N ,g^T\right) -1\right]
           \vec{g}^{\,T} -
           \left[\epsilon^N \left(g^N ,g^T\right)+1\right]\vec{g}^{\,N} \right\}
\end{eqnarray}
\label{dynamics}
\end{mathletters}  
\end{widetext}
and analogously for $\vec{v}^{\,\prime}_2$,
$\vec{\omega}^{\,\prime}_2$. With the use of eqs.~(\ref{boltz}) and
(\ref{dynamics}) and the above calculated restitution coefficients
$\epsilon^N \left(g^N ,g^T\right)$ and $~\epsilon^T \left(g^N ,g^T\right)$
(eqs.~(\ref{epsn}) and~(\ref{epst})) one can describe the evolution of
moderate dense granular gases without computing the detailed dynamics
of binary collisions as it is usually done in the ``soft sphere''
Molecular Dynamics (MD) approach. Here one considers the grains as
elastic bodies which deform each other during a collision. There are
several ansatzes for the force acting between touching
grains~\cite{CundallStrack,HaffWerner,Walton}. In all cases one has to
chose a time step for the integration scheme which is significantly
smaller than the typical collision time. Hence, during each collision one
has to calculate about 10\dots 1000 times the interaction force
between the grains to provide satisfying accuracy of the simulation.
When two grains approach each other they do not feel any interaction
as long as they do not touch each other. When granular particles
collide they interact via huge restoring forces which can be expressed
by Young moduli of the order of $Y=10^7~kg/m~sec^2$. The difficulty of
the simulation consists in the extreme short range interaction of the
particles and the resulting huge gradient of the interaction force.
Therefore presently one cannot simulate much more than 3000 granular
particles in three dimensions (e.g.~\cite{WaltonBraun,GallasEtAl}) and
about $10^4$ particles in two dimensions (e.g.~\cite{PoeschelBuchholtz}).

Another method for the simulation of granular assemblies is the ``hard
sphere'' approach where one does not consider the details of the
collision but only the pre-- and post--collisional velocities of each
pair of colliding grains. The goal of these simulations is the low
numerical complexity. One needs only computational effort when
particles collide but not in between the collisions. This allows for
the application of so called event--driven calculations
(e.g.~\cite{Rapaport}). Hence, one can simulate much more particles
than with ``soft particle'' methods.

One of the preconditions for the application of the ``hard sphere
approach'' is the exact knowledge of the normal and tangential
restitution coefficients, $\epsilon^N$ and $\epsilon^T$, as functions
of the normal and tangential impact rates, $\vec{g}^N$ and $\vec{g}^T$,
which theoretically to determine was the goal of the present paper.

An interesting possible application of this approach is the dynamics
of planetary rings composed of icy and silicate material which is
determined by inelastic dissipative collisions~\cite{Planetenbuch}.
The calculation of such systems using the traditional MD is impossible
due to the huge number of particles in these systems.

\section{Conclusion}
\label{conclusion}
A novel model for collision of particles in granular gases is
proposed. For the normal component of the relative motion the equation
of motion is derived based on the general consideration of the
viscoelastic impact. We find the expression for the dissipative part
of the normal force in the self--consistent quasistatic approximation
which generalizes the existing results for the viscoelastic
collisions~\cite{pao}. For the tangential relative motion we
investigated a microscopic model of surfaces of the
colliding particles which are in contact. We found a mean--field expression for the
tangential inter--particle force, which can reproduce smooth, reflecting or sticky collisions depending on the microscopic
parameters of the surfaces and on the relative impact velocity. A
frequently used model for collisions of granular particles by Haff and
Werner~\cite{CundallStrack} is contained in our model as a special
case. The restitution coefficients for the normal and tangential
motion are calculated as functions of the relative impact velocity. A
rather nontrivial strongly non--linear dependence of the tangential
restitution coefficient on the impact velocity is observed.

The obtained restitution coefficients may be used to describe the
dynamics of moderately dense granular gases, where the evolution
occurs via a sequence of successive binary collisions.

\acknowledgments We thank H.~J.~Herrmann and J.~T.~Jenkins for
discussion and S.~Luding for critical remarks. One of us (N.V.B.) is
very grateful to Prof. J.~Kurths for the warm hospitality in Potsdam.

\appendix
\section{Generalization of the Hertz theory}
\label{AppendixA}
We briefly sketch the Hertz theory of elastic impact
and give a generalization of this theory for the case of viscoelastic
collisions.

In the quasistatic approximation which is used in Hertz's impact
theory it is assumed that the (time dependent) strain and the
(time dependent) stress are related in the same manner as in the
static case. It may be shown (see appendix~\ref{AppendixB}) that this
approximation is valid for the elastic case when the characteristic
velocity is much less than the speed of sound in the material of the
colliding particles. Moreover for the viscoelastic case it is required
that the viscous relaxation time of the material is much shorter than
the duration of the collision.  In the static case the equation of
equilibrium reads~\cite{Landau}
\begin{equation}
\label{eqcond}
\nabla \cdot \hat{\sigma}_{(el)} = 0~,
\end{equation}
where the elastic stress tensor $\hat{\sigma}_{(el)}$ is expressed
in terms of displacements $\vec{u}(r)$ via
eq.~(\ref{stresstensor}). Hence the static eq.~(\ref{eqcond}) can be
written as
\begin{eqnarray}
&&\nabla^2_{\perp} \vec{u}+b^2 \, \nabla^2_{\parallel} \vec{u} = 
     0\label{static}\\
&&b^2=\frac{4E_1+6E_2}{3E_1}=\frac{2 \left( 1-\nu \right) }
     {\left( 1-2 \nu \right)}\nonumber
\end{eqnarray}
with the ``longitudinal'' and ``transversal'' parts of the Laplacian. 
\begin{mathletters}
\begin{eqnarray}
\nabla^2_{\parallel} &=& \nabla \circ \nabla\\
\nabla^2_{\perp} &=& \nabla^2- \nabla^2_{\parallel}~.
\end{eqnarray}
\end{mathletters}
The boundary conditions for the displacements in eq.~(\ref{static})
are formulated on the surface of contact.  From geometric
considerations it follows that the contact area between two colliding
particles is a plane. Using the appropriate coordinate system centered
in the middle of the contact region (where we set $z=0$) one can write
\begin{equation}
C_1x^2+C_2y^2+{u}_{z1}+{u}_{z2}=\xi~.
\end{equation}
The values ${u}_{z1}={u}_{z1}(x,y)$ and ${u}_{z2}={u}_{z2}(x,y)$ are
the $z$--components of the displacements in the materials of the
bodies at the plane $z=0$, $\xi $ is the total deformation (the sum of
the deformations of both bodies at the center of the contact area,
i.e. at $x=y=0$). The constants $C_1$ and $C_2$ are expressed in terms
of radii of curvature of the surfaces in contact (see
e.g.~\cite{Hertz,Landau}). The values of ${u}_{z1}$ and ${u}_{z2}$ may
be expressed in terms of the normal pressure $P_z(x,y)$ that acts
between the bodies at $z=0$~\cite{Landau}:
\begin{eqnarray}
{u}_{z1}(x,y,0) &=& {u}_{z2}=\frac{\Lambda}{\pi} 
          \int\int\frac{P_z\left(x^{\prime },y^{\prime }\right)}{r} 
          {d}x^{\prime }{d}y^{\prime }\\ 
r &=& \sqrt{\left(x-x^{\prime }\right)^2+\left(y-y^{\prime }\right)^2} \nonumber \\
\Lambda &=& \frac{ 2E_1+3E_2}{E_1\left( E_1+6E_2\right)}
         =\frac{1-\nu ^2}Y\nonumber~.
\end{eqnarray}
For simplicity we assume that the colliding particles are of the same
material. The normal pressure $P_z$ is related to the total normal
force $F_{(el)}$
\begin{equation}
P_z(x,y) = \frac{3~F_{(el)}}{2~\pi ab}
           \sqrt{1-\frac{x^2}{a^2}-\frac{y^2}{b^2}}~,
\end{equation}
where $a$ and $b$ are the semi--axes of the contact ellipse. The
latter values as well as the compression $\xi $ may be found from the
set of equations
\begin{mathletters}
\begin{eqnarray}
\xi &=& \frac{F_{(el)}}{\pi} \frac 32\Lambda \int^{\infty}_0 
      \frac{{d} q}{\sqrt{\left(a^2+q\right)\left(b^2+q\right) q }} \\
C_1 &=& \frac{F_{(el)}}{\pi}\frac 32\Lambda \int^{\infty}_0 
      \frac{{d} q}{\left(a^2+q\right)
      \sqrt{\left(a^2+q\right)\left(b^2+q\right) q}}\\
C_2 &=& \frac{F_{(el)}}{\pi} \frac 32\Lambda\int^{\infty}_0 
      \frac{{d} q}{\left(b^2+q\right)
      \sqrt{\left(a^2+q\right)\left(b^2+q\right) q}}~.
\end{eqnarray}
\end{mathletters}

From the above expressions follows that for all bodies in contact
having smooth surfaces (in the mathematical sense) the total force and
the deformation are related via the power law
\begin{equation}
F_{(el)}\left(\xi \right)=\tilde{c}\,\,\xi ^{3/2}~.
\end{equation}
The constant $\tilde{c}$ depends on the elastic properties of the
materials and on the local curvatures of the colliding bodies. For the
case of the spherical particles one has the Hertz's law 
\begin{equation}
F_{(el)}\left(\xi\right) = \frac{2~Y}{3~\left( 1-\nu ^2\right) }\sqrt{R^{eff}}\,
         \xi^{3/2}~.
\end{equation}
Using this relation between force and deformation and the equation of
motion (eq.~(\ref{gn1})) one can describe the elastic collision
completely. The duration of the collision is~(\cite{Hertz,Landau})
\begin{eqnarray}
t_c &=& 2.94\,\left(\frac{m^{\mbox{\it\footnotesize eff}}}
     {k}\right)^{2/5}\left( \left(g^N\right)\right) ^{-1/5}\label{duration}\\
k^2 &=& \left(\frac{4}{5}\frac{2}{3\,\Lambda}\right)^2 
     R^{\mbox{\it\footnotesize eff}}\nonumber~.
\end{eqnarray}

In the solution of the elastic contact problem the displacement fields
$\vec{u}_1\left(\vec{r}\right)$ and $\vec{u}_2\left(\vec{r}\right)$ are completely defined by the
value of $F_{(el)}$ and thus by the value of the deformation
$\xi$. Hence we write $\vec{u} \left(\vec{r}\right)= \vec{u} \left(\vec{r},\xi \right)$, i.e. the
displacement field depends explicitly on the compression. Therefore we
obtain for the velocity of the displacement in the quasistatic
approximation
\begin{equation}
\dot{\vec{u}}\left(\vec{r}\right) = \dot {\xi}\,\frac{\partial {}}{\partial {\xi} }
                    \vec{u}\left(\vec{r},\xi \right)
\end{equation}
and correspondingly for the dissipative part of the stress tensor
\begin{eqnarray}
\sigma^{ik}_{(dis)} &=& \dot{\xi}\,\,\frac{\partial {}}{\partial \xi}
     \left\{ \eta_1u_{ik} + 
     \left(\eta_2-\frac{\eta_1}3\right) u_{ll}\delta_{ik}\right\} 
     \nonumber \\ 
&=&\dot {\xi} \,\frac{\partial{}}{\partial \xi }\sigma^{ik}_{(el)}
     ~~~~~ ,~~~~
     \left(E_1\leftrightarrow \eta_1,\,E_2\leftrightarrow \eta_2 \right)~.
\end{eqnarray}
We emphasize that the expression in the curled brackets in the above
equation coincides with the elastic stress, provided the viscous
constants are substituted by the elastic ones. The latter expression
for the dissipative stress tensor is written for the case when the
memory effects in the viscous processes may be neglected. A more
general case is discussed in appendix~\ref{AppendixB}.

The $\sigma^{zz}_{(el)}$ component of the elastic stress is
equal to the normal pressure $P_z$ at the plane $z=0$, 
\begin{eqnarray}
&& \sigma^{zz}_{(el)}(x,y,0) =\nonumber \\
&& \hspace{1cm}= E_1\frac{\partial{u_z}}{\partial z} +
    \left( E_2-\frac{E_1}3\right) \left( 
    \frac{\partial {u_x}}{\partial x} +
    \frac{\partial{u_y}}{\partial y} + 
    \frac{\partial{u_z}}{\partial z} \right) \nonumber \\
&& \hspace{1cm}= \frac{3~F_{(el)}}{2~\pi ab}
    \sqrt{1-\frac{x^2}{a^2}-\frac{y^2}{b^2}}~.
\end{eqnarray}
With the transformation of the coordinate axes
\begin{mathletters}
\begin{eqnarray}
x& = & \alpha x' \\
y& = &\alpha y'  \\
z& = &z'
\end{eqnarray}
\end{mathletters}
and
\begin{mathletters}
\begin{eqnarray}
\alpha &=& \left(\frac{\eta_2-\frac 13\eta_1}
         {\eta_2+\frac 23\eta_1}\right) 
         \left(\frac{E_2+\frac 23E_1}{E_2-\frac 13E_1}\right)\\
\beta &=& \frac{\left(\eta_2-\frac 13\eta_1\right)}
         {\alpha\left(E_2-\frac 13E_1\right) }\\
a&=&\alpha a^{\prime}\\
b&=&\alpha b^{\prime}
\end{eqnarray}
\end{mathletters}
we obtain
\begin{eqnarray}
&& \eta_1\frac{\partial {u_z}}{\partial z} +
   \left(\eta_2-\frac{\eta_1} 3\right)\!\!\left( 
   \frac{\partial {u_x}}{\partial x} +
   \frac{\partial {u_y}}{\partial y} +
   \frac{\partial {u_z}}{\partial z} \right) \nonumber \\
&=& \beta \!\!\left( E_1\frac{\partial{u_z}}{\partial z^{\prime}}+
   \left(E_2-\frac{E_1}3\right)\!\! \left( 
   \frac{\partial {u_x}}{\partial x^{\prime }} +
   \frac{\partial {u_y}}{\partial y^{\prime }} +
   \frac{\partial {u_z}}{\partial z^{\prime }}\right)\! \right) 
   \nonumber\\
&=& \beta \frac{3~F_{(el)}}{2~\pi a^{\prime}b^{\prime}}
   \sqrt{1-\frac{x^{\prime}{}^2}{a^{\prime}{}^2} -
           \frac{y^{\prime}{}^2}{b^{\prime}{}^2}}\nonumber \\
&=&\beta \alpha ^2 \,\frac{3~F_{(el)}}{2~\pi ab}
   \sqrt{1-\frac{x^2}{a^2}-\frac{y^2}{b^2}}~.
\end{eqnarray}
Applying the operator $\dot {\xi }\,\frac{\partial {}}{\partial \xi }$
to the previous expression we obtain the result for the viscous
stress. Integrating the viscous stress over the contact area we
finally find for the dissipative component of the inter--particles
force
\begin{eqnarray}
&&F_{(dis)} = A\,\dot{\xi}\frac{\partial{}}{\partial\xi}F_{(el)} \left(\xi \right)\\
&&A = \alpha ^2\beta = \frac 13 
   \frac{\left(3\eta_2-\eta_1\right)^2}
        {\left( 3\eta_2+2\eta_1\right)}
   \frac{\left(1-\nu ^2\right)\left(1-2~\nu \right)}
        {Y\,\nu^2}~.
\end{eqnarray}
Thus one obtains for the normal force which acts between the
viscoelastic bodies in the quasistatic regime of collision
\begin{equation}
F = const\,\left( \xi ^{3/2} +
    \frac 32A\, \sqrt{\xi} ~\dot {\xi} \right)~.
\label{43}
\end{equation}
The constant in eq.~(\ref{43}) coincides with that for the elastic
force. For colliding spherical particles we arrive at
eq.~(\ref{forcedis}).

The impact theory, sketched above, was developed for the bodies with 
the smooth surfaces. If the surface asperities are taken into account,
one can consider the actual surface as a smooth one (obtained by 
averaging over the asperities heights),  with a small perturbation 
superimposed due to the presence of the asperities.
One can also consider the actual normal
displacements and normal pressure as a sum of the averaged 
(over the asperities) values and the small perturbation. Then it is easy 
to show that the equations, obtained for the averaged values coincide 
(due to linearity of the problem) with the corresponding equations for 
the elastic collision of the smooth bodies. As a result, the relation 
between the force and deformation, $\xi$  is the same as in the Hertz's 
theory, provided that $\xi$ is defined with a use of the averaged over 
the asperities radii of the colliders. 

Considering the normal motion for the dissipative collisions, one need
not to consider the plastic deformation of the asperities, since the
size of the asperities is assumed to be very small compared with the
radii of the spheres. For our calculations in fig.~2. the asperity
size is $10^3$ to $10^7$ times smaller than the effective Radius of
the particles. Hence the dissipation in the bulk of the asperities is
negligibly small compared to the total dissipation in the compressed
part of the collider. Moreover, the ratio of the normal to tangential
stress, may be roughly estimated as,
$\overline{\sigma^N}/\overline{\sigma^T} \sim \left(\xi /R
\right)^{1/2}$, so that the crushing of the asperities does not seem
to be important for the normal motion, if $\xi /R \ll 1$ and if the
conditions of the quasistatic collision hold. Thus one concludes, that
the surface asperities may be ignored, when the normal motion is
studied, provided they are small and the conditions of the quasistatic
collision are satisfied.

\section{Validity of the quasistatic approximation}
\label{AppendixB}
To analyze more rigorously the conditions when the quasistatic
approximation is valid we write the equation of motion for the
viscoelastic continuum
\begin{equation}
\label{geneqmot}
\rho\ddot{\vec{u}} = \nabla \cdot \left(\hat{\sigma}_{(el)} +
   \hat{\sigma}_{(dis)} \right)
\end{equation}
where $\rho$ is the density of the material. The expression for the
elastic part of the stress tensor is given by
eq.~(\ref{stresstensor}). Taking into account the memory effects of
the dissipative processes in the material one can write for the
dissipative part
\begin{widetext}
\begin{equation}
\label{dismem}
\hat{\sigma}_{(dis)} \left( t \right) = E_1 \int_0^t \psi_1 
   \left(t-\tau \right) \left[\frac{1}{2}\left\{
   \nabla\circ \dot {\vec{u}}(\tau) + 
   \dot{\vec{u}}(\tau) \circ\nabla\right\} -
   \frac{1}{3}\hat{I}\nabla\cdot\dot{\vec{u}}(\tau)\right]
+ E_2 \int_0^t \psi_2 \left( t-\tau \right)\hat{I}\, \nabla \cdot 
   \dot{\vec{u}}(\tau)
\end{equation}  
\end{widetext}
where the (dimensionless) functions $\psi_1(t)$ and $\psi_2(t)$ are
relaxation (or ``memory'') functions for the distortion strain and the
dilatation, respectively. Note that eq.~(\ref{dismem}) coincides with
the corresponding expression for the viscous stress tensor
in~\cite{pao,impact} for $\psi_2(t)=0$. The latter approximation means
that one neglects the bulk dissipation due to the dilatation rate. For
the normal motion of colliding particles, however, the dissipation of
energy due to the dilatation rate and the dissipation due to the
distortion strain rate are of the same order of magnitude. Thus we
keep both relaxation functions in our considerations. Introducing
transversal and longitudinal velocities of sound in the material
\begin{mathletters}
\begin{eqnarray}
c_t^2 &=& \frac{E_1}{2\,\rho}=\frac{Y}{2\, \rho \left(1+ \nu\right)}\\
c_l^2 &=& \frac{2E_1 + 3 E_2}{3\, \rho}=\frac{Y\left(1-\nu\right)}{\rho\,
        \left(1+\nu\right)\left(1-2\nu\right)}\\
b^2 &=& \frac{c_l^2}{c_t^2}= \frac{2\left(1-\nu\right)}{\left(1-2\nu\right)}\nonumber
\end{eqnarray}
\end{mathletters}
one can write the equation of motion for the viscoelastic medium
\begin{widetext}
\begin{equation}
\label{viscgeneral}
\frac{1}{c_t^2} \ddot{\vec{u}} = \left\{\nabla^2_{\perp}\vec{u} +
   b^2\,\nabla^2_{\parallel}\vec{u}\right\}+\nabla^2_{\perp}\psi_1
   \ast \dot{\vec{u}} +
 \nabla^2_{\parallel}\left\{ \frac{4}{3}\psi_1\ast\dot{\vec{u}} +
   \left(b^2 -\frac{4}{3} \right)\psi_2 \ast \dot{\vec{u}} \right\} 
\end{equation}
  \end{widetext}
where $\psi \ast \dot{\vec{u}}$ denotes convolution.

To estimate the relative importance of the terms in
eq.~(\ref{viscgeneral}) we introduce the characteristic velocity
$v_0=g^N$ and the characteristic time $\tau_0=t_c$, where $t_c$ is the
duration of the collision, introduced above in eq.~(\ref{duration}).
Then the characteristic length is $R_0=v_0~
\tau_0$. Eq.~(\ref{viscgeneral}) can be written in a dimensionless form
\begin{widetext}
\begin{equation}
\label{viscgen1}
\left(\frac{v_0^2}{c_t^2}\right)\ddot{\vec{\tilde{u}}}
\left(t\right) = 
   \left\{\nabla^2_{\perp}\vec{\tilde{u}} \left( t \right) +
   b^2\,\nabla^2_{\parallel}\vec{\tilde{u}}\left(t\right)\right\}+
   \left(\frac{\tau_{vis,1}}{\tau_0} \right) 
   \nabla^2_{\perp}\dot{\vec{\tilde{u}}}\left(\tilde{t}_*\right) + 
\left( \frac{\tau_{vis,1}}{\tau_0} \right) \nabla^2_{\parallel}
   \left\{\frac{4}{3}\dot{\vec{\tilde{u}}}\left(\tilde{t}_*\right)+
   \left(b^2-\frac{4}{3}\right) 
   \left(\frac{\tau_{vis,2}}{\tau_{vis,1}}\right)
   \dot{\vec{\tilde{u}}} \left( \tilde{t}_* \right) \right\}
\end{equation}  
\end{widetext}
\begin{equation}
\tau_{vis,1/2} = \int_0^t \psi_{1/2} \left( \tau \right) d \tau~.
\end{equation}
We use the following representation of the convolution
\begin{equation}
\psi\ast\dot{\vec{u}} = \dot{\vec{\tilde{u}}}\left(\tilde{t}_*\right) 
   ~\int_0^t \psi_{1/2}\tau d \tau 
\label{convolution}
\end{equation}
with $\tilde{t}_* $ being a dimensionless time from the interval
$0\leq \tilde{t}_*\leq \tilde{t} = t/\tau_0 $. The relation for the
convolution (\ref{convolution}) is valid if $\psi(t) \ge 0$.

During the collision process $t$ is of the order of $\tau_0$,
i.e. $\tilde{t}$ is of the order of $1$, while by the definition of
$\tau_{vis,1/2}$ these values are of the order of the relaxation
times for the dissipative processes in the material. That means, that
$\tau_{vis,1/2}$ characterizes the time when the memory effects are
important. If the duration of the collision is much greater than the
relaxation times, i.e. if $\tau_{vis,1/2} \ll \tau_0 $, one can write
\begin{equation}
\label{reltimes}
\tau_{vis,1/2}\simeq\int^{\infty}_0\psi_{1/2}\left(\tau\right) d \tau
\end{equation}
and consequently
\begin{equation}
\psi\ast\dot{\vec{u}}\simeq\dot{\vec{u}}\left(t\right)\tau_{vis,1/2}~. 
\end{equation}
If the characteristic velocity $v_0$ is much less then the speed of
sound in the material too, one can neglect the terms with vanishing
factors $\left( v_0^2/c_t^2 \right)$ and $\left(
\frac{\tau_{vis,1/2}}{\tau_0} \right)$ in eq.~(\ref{viscgen1}) and
finally one arrives at the static eq.~(\ref{static}). That means the
quasistatic approach is valid provided that the conditions
\begin{mathletters}
\begin{eqnarray}
1 &\gg& \frac{v_0^2}{c_t^2} \\
1 &\gg& \frac{\tau_{vis,1/2}}{\tau_0}
\end{eqnarray}
\end{mathletters}
hold. From the above considerations follows that in the quasistatic
approximation the memory effects in the dissipative processes are not
important and the viscous part of the stress tensor may be written in
the same way as in eq.~(\ref{stresstensor}), with the viscous
constants $\eta_1$ and $ \eta_2$ given by
\begin{equation}
\eta_{1/2} = E_{1/2}~ \tau_{vis,1/2} 
           = E_{1/2}~ \int^{\infty}_0\psi_{1/2}
             \left( \tau \right) d \tau
\end{equation}

It is worth to note that the quasistatic approximation is valid for
many of the granular gases one encounters in nature, since usually the
characteristic velocity in these systems is low. One should also note
that the description of the collision in the quasistatic approximation
is rigorous in a sense that no other additional approximations are
used.

As it follows from the above considerations, it is not correct to use
the time dependent relaxation functions for the dissipative part of
the stress tensor together with Hertz's quasistatic
relations~\cite{pao,impact}, since this approach is not
self--consistent and one has to assume a lot of additional hardly
controllable approximations.

\end{document}